# THE effect of the number OF Mobile Nodes on varying speeds of MANETS

John Tengviel[1], K. Diawuo[2] and K. A. Dotche[3]

[1]Department of Computer Science, Sunyani Polytechnic, Sunyani, Ghana

john2001gh@yahoo.com

[2]Department of Computer Engineering, KNUST, Kumasi, Ghana

kdiawuo@yahoo.com

[3]Department of Telecommunications Engineering, KNUST, Kumasi, Ghana

kdotche2004@gmail.com

*Abstract*

*Mobile Ad hoc Networks (MANETs) are dynamic networks populated by mobile devices, or mobile nodes (MNs).The Mobile Nodes (MNs) are free to move anywhere and at any time*. *The population of the nodes may have some influence on the mobility rate of the MNs. This paper presents simulation results using MatLab software. The study investigates the influence of mobile nodes' parameters such as number of nodes on the nodes speeds and nodes distribution in a given area. The results have indicated that the number of mobile nodes have impact on the speeds of the nodes in a location.*

## 1. Introduction

Mobile Adhoc NETworks (*MANETs)* is a collection of wireless mobile nodes configured to communicate amongst each other without the aid of an existing infrastructure. MANETS are *Multi-Hop* wireless networks since one node may not be indirect communication range of other node. Ad hoc networks are viewed to be suitable for all situations in which a temporary communication is desired. The technology was initially developed keeping in mind the military applications [1, 2] such as battle field in an unknown territory where an infrastructured network is almost impossible to have or maintain. In such situations, the ad hoc networks having self-organizing [3-6] capability can be effectively used where other technologies either fail or cannot be effectively deployed. The entire network is mobile, and the individual terminals are allowed to move freely. Since, the nodes are mobile; the network topology is thus dynamic. This leads to frequent and unpredictable connectivity changes. In this dynamic topology, some pairs of terminals may not be able to communicate directly with each other and have to rely on some other terminals so that the messages are being delivered to their destinations. Such networks are often referred to as multi-hops or store-and-forward networks [1 ]. An illustration is shown in Figure 1.

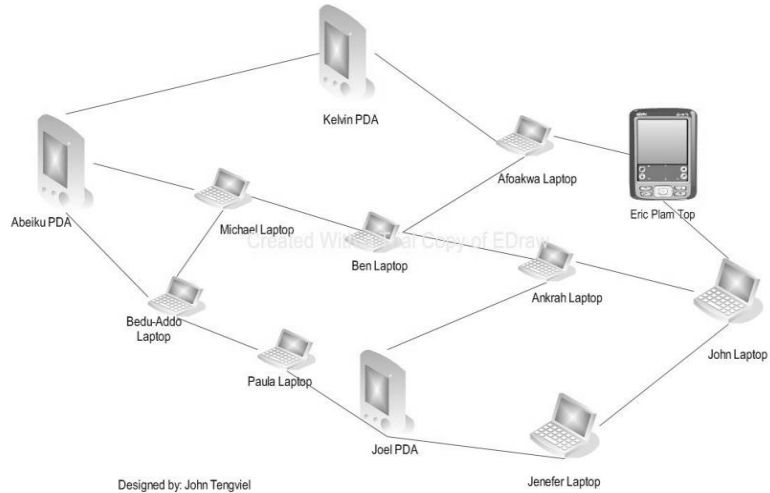

**Figure 1: MSC Telecommunication Engineering Students Sharing Information**

Figures 1 show an example of an ad hoc network which has different communication devices and some connections amongst them; in which the MSC *students were to do a group work/discussion on campus.* The students come in the area and without the need of any existing infrastructure, switch on their handsets, laptops or PDAs that enable them to communicate with each other while moving and carrying out their work from one office to another office or one floor to another. In this case of group work, an ad hoc network has been formed by the communication devices.

This paper presents a study on mobile nodes in MANETs using Mobiles' Node Speed Correlation. The section 2 illustrates a brief review on MANETs studies. The section 3 introduces the Node Speed Correlation and the ratios of the nodes speeds to the number of nodes. The simulation procedures and considered parameters are presented in section 4. The obtained results are observed and analysed in section 5. Finally section 6 concludes the paper and the direction we would like to take this work in the future.

## 1.1. Goals of the Paper
The aim of the paper was to study the impact of mobile nodes' parameters such as the speed and the size of mobile nodes in a given area using Node Speed Correlation. The following areas were being explored:

- An analysis of mobility model parameters using the statistical distributions.
- A development of a set of metrics for node mobility.
- Analyse the performance with MatLab simulation

## 2. RELATED WORKS
The study of Ad hoc network seem to be centered around the user mobility. Mobility models are mainly used to describe the movement of nodes, so they play a key part in simulating ad hoc networks [1, 7]. Nodes operate in real-life situations. The user or any mobile device is modeled as a node.

## 2.1. Mobility Models For Ad Hoc Networks

The MANETs models' study in [1-6], has shown that currently there are two types of mobility models used in simulation of networks. These are traces and synthetic models. Traces are those mobility patterns that are observed in real-life systems. Traces provide accurate information, especially when they involve a large number of mobile nodes (MNs) and appropriate long observation period. On the other hand, synthetic models attempt to realistically represent the behaviour of MNs without the use of traces. They are divided into two categories, entity mobility models and group mobility models [1, 4, 6, 7, 8]. The entity mobility models randomise the movements of each individual node and represent MNs whose movements are independent of each other.

### 2.1.1. Random Walk Mobility

The Random Mobility Model (RMM) for ad'hoc networks is the Random Walk/ Brownian Mobility Model used in cellular networks. In this modeling the current speed and direction of a Mobile Node (MN) is independent of its past speed and direction [1, 3, 8]. Thus, we encounter an unrealistic generation of movements such as sudden stopping, sharp turning, and completely random wandering. Due to these difficulties, many authors modify the Random Mobility Model by changing the calculation of speed, direction, or both. The random walk mobility (RWM) model is defined by Camp et al [7], as an erratic movement. This might appear quite very difficult to predict. It is a memoryless movement whose previous speeds and directions are unknown. This property produces a simple mathematical mobility model which makes use of Markov processes and chains, with a generalised unrealistic approach in user's movement. The node position is modeled with a random speed (V) and direction (θ). Both of them are chosen from predefined ranges. The speed is defined in the range of a minimum speed (Vmin) and maximum speed (Vmax) and written as [Vmin,Vmax] having a direction [0, 2π]. Each movement has a constant duration or a constant distance travelled.

### 2.1.2. Random Waypoint Mobility Model

The Random Waypoint Mobility (RWM) includes pause times between changes in direction and/or speed [ 4, 6, 8]. MN begins by staying in one location for a certain period of time known as a pause time which is not available in the previous models. Once this time expires, the MN chooses a random destination with a speed that is uniformly distributed over the range [0, Vmax]. It travels towards the newly chosen destination at the selected speed. Upon arrival, the MN takes another break (pause) before starting the process again. Many authors [7] adopted this model in their simulation studies including slightly modified Random Waypoint Mobility Model so that MN travels at a constant speed throughout the entire simulation. This model is a memoryless, and has the same limitation as the random mobility model.

### 2.1.3 Probabilistic Version of the Random Mobility Model

In [4, 6, 8], the mobility model utilizes a probability matrix to determine the position of a particular MN in the next time step. The probability matrix takes into account three different states. These states are: State 0 represents the current location of a given MN, state 1 represents the MN's previous location, and state 2 represents the MN's next location if the MN move forward. The probability matrix is written as:

$$P = \begin{bmatrix} P(0,0) & P(0,1) & P(0,2) \\ P(1,0) & P(1,1) & P(1,2) \\ P(2,0) & P(2,1) & P(2,2) \end{bmatrix}$$

Where each entry P(a,b) represents the probability that an MN will go from **state A** to **state B**. In [7, 9], Chiang's simulator uses each node that moves randomly with a preset average speed. The values use to calculate x and y movements are given as:

$$P_1 = \begin{bmatrix} 0 & 0.5 & 0.5 \\ 0.3 & 0.7 & 0 \\ 0.3 & 0 & 0.7 \end{bmatrix}$$

Where, x is the horizontal movement and y is the vertical movement.
Probability matrix **P1** allows an MN to move in any direction as long as it does not return to its previous position. This implementation produces a probabilistic movement rather than purely random movements, thus yields more realistic behaviour. For example, as people complete their daily tasks they tend to continue moving in a semi-constant forward direction. Authors in [2] have argued that rarely we do unexpected turn around to retrace our steps and almost never take random steps hoping that we eventually wind up somewhere relevant to our tasks. However, choosing appropriate values of P($a$, $b$) may seem difficult, if not impossible.

## 2.2. Group Mobility Model
Group mobility models represent multiple MNs having a total or partial action dependent of one another [2]. However, in many real situations, it is necessary to model the behavior of MNs that move together. For example, many military scenarios occur where a group of soldiers must collectively search a particular plot of land in order to destroy land mines, capture enemy attackers, or simply work together in a cooperative manner to accomplish a common goal [4, 6, 7, 9, 10]. In such situations, group mobility model need to account for new cooperative characteristics.

### 2.2.1. Simple Group Mobility Models by Sanchez

Sanchez et al [10] have noted that a random walk/random movement model may not be sufficient to describe many "real-life" situations. Typically, the model may account for dependencies resulting from the interactions between MNs. These interactions have been considered in Pursue Model, and Nomadic Community Mobility Model [4, 5, 6, 9, 10].

### 2.2.2. Pursue Mobility Model

The Pursue Mobility Model is also defined in [10, 11, 12]. As the name implies, the Pursue Mobility Model attempts to represent MNs tracking a particular target. For example, this model could represent police officers attempting to catch an escaped criminal. The Pursue Mobility Model consists of a single update equation for the new position of each MN:
*new position = old position + acceleration(target - old position) + random vector*
where *acceleration(target - old position)* is information on the movement of the MN being pursued and *random vector* is a random offset for each MN. The *random vector* value is obtained via an entity mobility model (e.g., the Random Walk Mobility Model); the amount of randomness for each MN is limited in order to maintain effective tracking of the MN being pursued. The current position of an MN, a random vector, and an acceleration function are combined to calculate the next position of the MN.

## 3.0 Model of Study

### 3.1. Mobiles' Node Speed Correlation

A metric of investigation node mobility is the speed correlation $Cor_{MNspeed}$. For a given different time $t$ and $t + \Delta t$ a temporal dependence degree is assumed between two mobiles nodes in motion with speed $\vec{v_1}$ and $\vec{v_2}$. The speed correlation $Cor_{MNspeed}$ can be expressed as in [13] and given by equation:

$$Cor_{MNspeed} = \frac{\vec{v_1} * \vec{v_2}}{\|\vec{v_1}\| * \|\vec{v_2}\|} * \frac{\|\vec{v_1}(t+\Delta t)\|}{\|\vec{v_2}\|} \qquad 1$$

It may also be admitted that the speed correlation is not effective to describe node mobility modelling, since the movement of nodes is not uniform over a given area as well as in simulation works. It is suggested in [13] that a reliable mobility node model should minimizes the distance between nodes. We also recall a clustering coefficient of nodes and assume that as in real-world, group has impact on individuals. The clustering coefficient as proposed by Watts & Stogatz is a metric used to discriminate mobility models. The clustering coefficient is the ratio of the radio links among neighbours and the number of neighbours' nodes. This clustering coefficient is related to the network redundancy and is calculated as: Let $N_{MN}$ be the set of neighbors of the node $MN$ and $N_v$ the total number of nodes. The clustering coefficient $Clust_{coeff}$ is defined as:

$$Clust_{coeff} = \frac{\sum_{v \in N_{MN}} |\{x \in N_v\}|}{N_{MN}} \qquad 2$$

Consequently, $Clust_{coeff}$ states that if a node presents a high clustering coefficient, this means that the neighboured nodes have more radio links. The control over the clustering coefficient and node speed correlation may highly improve the routing performance but though this is not the concern in the study of modeling mobile node mobility. In this work, we assume that over a given area with following hypothesis:

If we let $d$ be the maximum distance between Nodes, let $\Delta x$ be a dimensionless displacement which corresponds to time difference $\Delta t$ such that:

If $\Delta x < d$ then $\lambda$ may be assumed to follow Poisson arrival rate

If $\Delta x > d$, the MN moves out of the given area then $\lambda$ may be assumed to follow Pareto arrival rate [13].

Hence, we can define a factor of correlation distance $Cor_{dist}$ between nodes as:

$$Cor_{dist} = k * \sum \frac{\|d(t), \ d(t+\Delta t)\|}{\|d\|} \qquad 3$$

K is a parameter which is related to the total number of nodes and the node speed of arrival.

### 4. METHODOLOGY

Our objective is to find out the influence of the number of mobile nodes on varying speeds of MNs and the speed correlation.

### 4.1. Impact of Varying Number of Nodes on Speeds

Various speeds against varying nodes densities were being assumed. Speed has been set to 1m/s by truncating Gaussian (normal) mean of human walking speed of 1.34m/s. This has been accomplished by placing upper and lower bounds for the mobility rate and the speed used in the simulation range from 0.1m/s to 1m/s. with 0.1 step.

The number of nodes in the network was varied and their effect on speed were considered because network size in combination with simulation area provides a good measure of the density of network. The mobility algorithms' performance has been evaluated for cases where nodes' density is assumed to be dense or sparse. In this scenario, the number of nodes has been varied from 50 to 300.

**Table 1. Simulation Parameters**

| Parameters | Values |
|---|---|
| Number of Nodes | 50, 100, 150, 200, 300 |
| Speeds | 0.1m/s to1m/s |

The simulation was done through MatLab software [1, 14].

## 5. RESULTS AND DISCUSSION

## 5.1. Effect of Varying Number of Nodes on Speeds

The Figure 2 shows the results of speeds ratios varying density of nodes within the network area node speed correlation. Simulation results on MATLab exhibit the effect of the MNs population on the mobility rate.

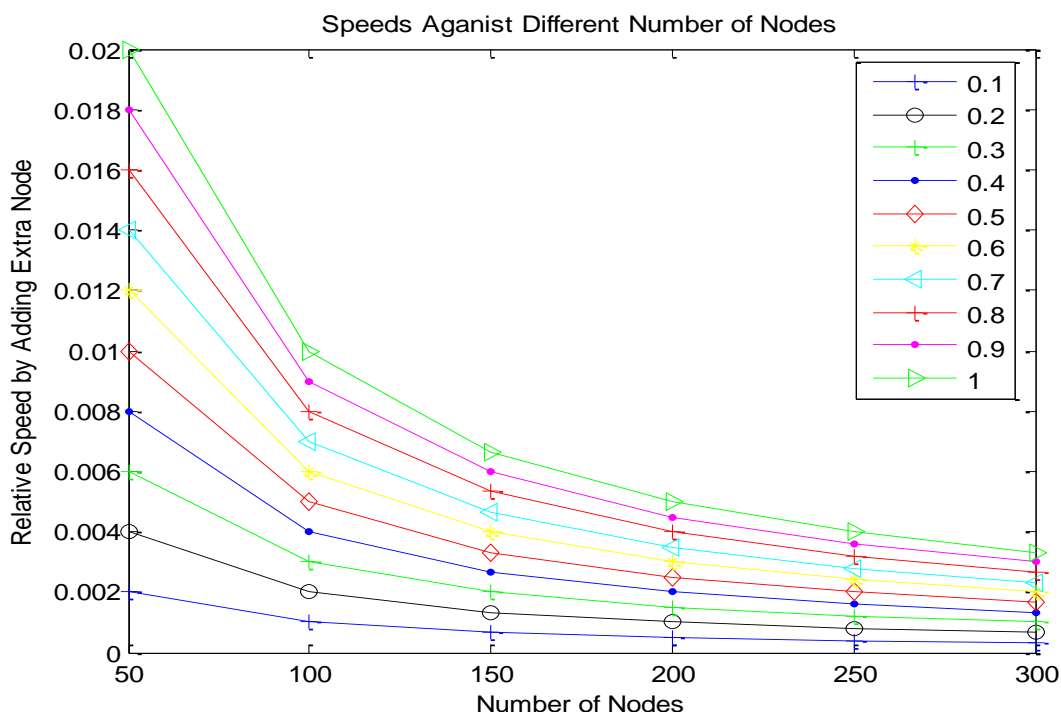

**Figure 2. Variation of MNs on Different Speeds**

The observation analysis may indicate that speeds ratio decreases with increases in the number of MNs within the network. The physical implication of mobility rate in node encounters has been glossed over. In reality, the total connection time of a node over a specific interval depends on the nodes encounter rate and the time in each encounter, both of which depend on the relative mobility of nodes [1]. Although a high node speed results in more node encounters, the connection time in each node encounter also decreases. When a node with a higher desired speed catches up with a slower moving node, it will either follow or overtake (bypass) [16]. The impact of this relationship is that nodes can and will be tightly packed resulting to High density, if their speed is low resulting to congestion, but if the speed is higher, the nodes would be farther apart resulting to low density. In Figure 2 illustrates that one need not have many nodes before a slow node becomes a bottleneck.

In Figure 2 it can be observed that as the number of the nodes increases, the speed of the nodes decreases. This is because nodes are closed to each other making mobility difficult. One may say that the network under study is crowded with nodes. As the number of nodes increases it could be better to increase the speed of the nodes, so that the nodes can move fast to give room to other nodes. In all, if the number of nodes is higher then speed must be increased for better mobility which is the opposite in the case of fewer nodes. It also means that nodes have a number of hops to get to their destination nodes. The larger the number of nodes means it require higher speed in order to get to a particular location. A decreased in the number of nodes in an area implies a decreased in the connectivity of nodes i.e., each node has fewer neighbours. A decreased in connectivity also implies lesser information exchange hence less input to the algorithm. An increased in the number of nodes implies high connectivity among nodes; more information is exchanged and hence more input to the algorithm. It is therefore important to conclude that when the nodes are many in a particular location, it would be wise to increase the speed to a certain limit. It could be observed that in all cases of speed against increased in nodes there was a decreased in speed. The distributions of speeds for the nodes have an impact on their connectivity and mobility. However, a large span of speeds also results in a wide range of connectivity.

## 5.2 K-Factor for Distance Correlation on Mobile Node Speed

In order to determine the k factor as given in equation 3 and given as equation 4:

$$Cor_{dist} = k * \sum \frac{\|d(t),\ d(t+\Delta t)\|}{\|d\|} \qquad 4$$

We applied Kolmogorov-Smirnov (KS) test [15] to Figure 2 which yielded results in Table 2. The result in Table 2 was used to plot the $k$-factor against the number of mobile nodes as shown in Figure 3.

$$k = |\text{ymax} - \text{ymin}| \qquad 5$$

**Table 2. Relative Difference between Maximum and Minimum Speeds Against Number of Nodes.**

| MN | Ymin | Ymax | $k$ |
|---|---|---|---|
| 50 | 0.002 | 0.02 | 0.018 |
| 100 | 0.001 | 0.01 | 0.009 |
| 150 | 0.00067 | 0.0067 | 0.00603 |
| 200 | 0.0005 | 0.005 | 0.0045 |
| 250 | 0.0004 | 0.004 | 0.0036 |
| 300 | 0.00033 | 0.0033 | 0.00297 |

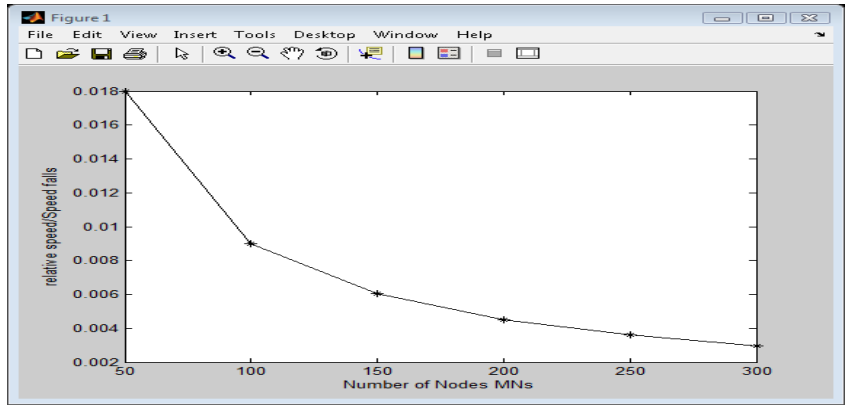

**Figure 3: Relative fall as the number of nodes increased**

It may also be observed that the relationship between the k values and the number of nodes is an exponential function with a negative decay [16, 20].

Therefore the distance correlation as in Equations 3 and 4 may be approximated to one Pareto shape [20] factor function Mean denoted as $\alpha_1 = 0.735$ which the mean value of k is. It comes that:

$$f(x) = \left(\frac{0.735}{1+X}\right)^{0.735}$$

With regard to $\alpha_2 = 0.527$ as the median value of k, we can write $f(x) = \left(\frac{0.527}{1+X}\right)^{0.527}$.

A comparative study is made in Figure 4 which depicts the use of the mean and median as one shape factor. Obviously, both quantities give the same layout meanwhile one may argue that the use of the median gives a better approximation as obtained in the Figure 4. Thus it may indicate that the distance correlation may be achieved if one considers the median value of the distance correlation parameter for a better approximation in mobility prediction for geocasting network [17] and may certainly be helpful for location awareness system as needed to characterize human mobility [17, 18].

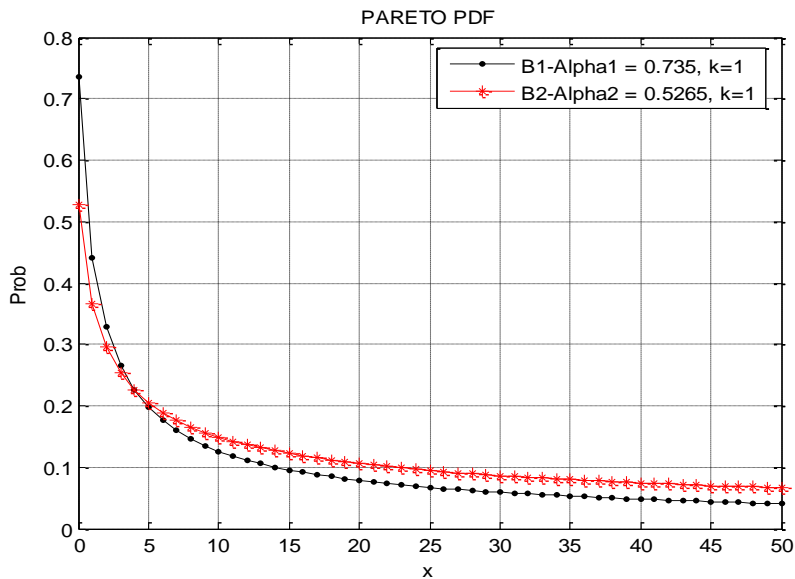

**Figure 4: Single Parameter Pareto Distribution with Fifty Nodes**

Over the years, it turns out that the speed ratio node density was always a weak point in design of mobility model. But now we can find that the distribution of speeds over the number of nodes fits heavy-tail/Pareto distribution according to our result. The distribution of speeds ratio number of nodes has been proved to be realistic, since the relation we used to connect speeds and number of nodes is also realistic [19, 20].

## 6. CONCLUSION

We have shown the effect of the number of mobile nodes the speeds of the MNs and their distribution in a location. It may claim that as the number of MNs increased the speeds of MNs may also fall but to a certain limit. It was therefore necessary to increase the speed MNs to give room to other nodes or make it possible for free movement.

Furthermore, the simulation results based on number of nodes to speeds relationship indicated that a better mobility can be achieved with higher speed as the number of nodes increased and performed better with a small number of nodes move with lower speed. Our future work will be to investigate how this would work in real-life situation.

**Authors**

**John Tengviel**

He is a holder of a BSc. Computer Science from University of Science and Technology (KNUST) in 2001 and currently candidate of MSc. Telecommunication Engineering from College of Engineering at the same university. He is a senior instructor with the Department of Computer Science at Sunyani Polytechnic. His research interests include Mobile Ad hoc Networks and Mobility modeling in MANETs.

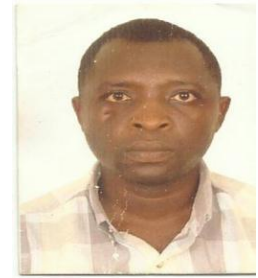

**Nana (Dr.) Kwasi Diawuo**

He is a senior lecturer of the Department of Computer Engineering at Kwame Nkrumah University of Science and Technology (KNUST), Kumasi, Ghana. He earned a BSc. (Electrical/ Electronic Engineering) from KNUST, M.Sc., Ph.D, MGhIE. He is a member of the Institution of Electrical and Electronic Engineers (IEEE) and Computer Society (of IEEE).

**K. A. Dotche**

He is a holder of a BSc. Electrical Eng. from University of Lome and MSc. Telecom. Eng. from College of Engineering at Kwame Nkrumah University of Science and Technology (KNUST); respectively 2004 and MSc 2010. He is currently a Ph. D. research candidate with the Department of Telecommunications Engineering, at KNUST. His research interests include Energy efficiency in wireless sensor networks, antennas and E-M propagation in cellular layered networks.

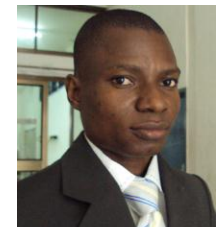